# The First Principle of Thermodynamics and the Non-Separability of the Quantities "Work" and "Heat": The adiabatic piston controversy

Rodrigo de Abreu
Centro de Electrodinâmica e Departamento de Física do IST, Lisboa, Portugal

Abstract

The impossibility of separating into work and heat the energy transmitted between two subsystems through a movable piston is analyzed in this article. The process here described, although "quasi-static", is not reversible. It is shown that the First Principle, $dU=dW+dQ$, introduced by Clausius, does not generally allow a physical identification of $dW$ and $dQ$, although $dU=-pdV+TdS$ is verified along the equilibrium points of each subsystem.

## 1 - Introduction

The concepts of work and heat appear associated in the First principle of Thermodynamics, $dU=dW+dQ$. Although concepts such as work and heat historically originated prior to Clausiu´s work [1, 2], the formalism emerging from the First Principle culminates with the introduction of the entropy through the expression $dS=dQ/T$. The success of Clausius's analysis explains the subsequent reinterpretation of the quantities work and heat involved [3] in the First principle. This reinterpretation imposed itself by its formal success, the acceptance of which is almost universal today. The contradictions [4, 5] of these reinterpretations have been analyzed elsewhere [6]. In this paper one simple but fundamental aspect is analyzed.

## 2 - The First Principle of Thermodynamics, the energy conservation principle and the concepts of work and heat

Consider the situation represented in Fig.1. Assume that on the sides 1 and 2 of the piston there are atoms of the same monatomic ideal gas. The piston is made of a material for which the thermal conductivity is zero and the piston is initially blocked so that it cannot move. Thus, initially no energy can pass from side 1 to side 2. We take the initial temperatures, $T_1$ and $T_2$, to be different, and the initial pressures, $p_1$ and $p_2$, to be equal. Once unblocked, the piston gains a translational energy to the right of order $1/2KT_1$ from a collision with a side 1 molecule, and a translational energy to the left of order $1/2KT_2$ from a collision with a side 2 molecule [7-37]. In this way energy passes mainly from side 2 to side 1 if $T_2>T_1$.
Generally if the thermal conductivity is not zero energy can pass through the piston. But in the ideal case we are considering, the energy can only pass if the piston is moving.

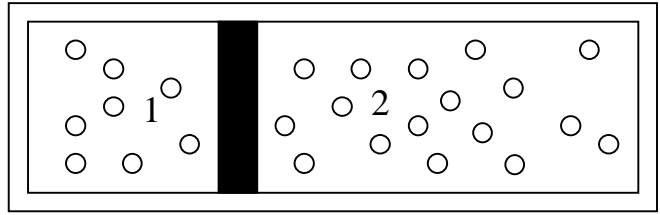

Fig. 1 An adiabatic piston surrounded by a gas. The initial values of pressures $p_1$ and $p_2$ are equals. The initial temperatures $T_1$ and $T_2$ are differents. The mass of the piston is large compared with the mass of the particles and the piston moves, jiggling, to an equilibrium point with $T_1 = T_2$.

Since in the case the piston is "adiabatic" ("$dQ$"$=0$ for the piston) and the process is "quasi-static" (there is equilibrium at all times on sides 1 and 2 and the pressure and temperature are well defined on each side), one might anticipate that the process is also reversible and thus isentropic (adiabatic reversible). In fact this is not true, as can be confirmed by solving the isentropic equations,

$$p_1 V_1^\gamma = p_1' V_1'^\gamma \qquad (1)$$

and

$$p_2 V_2^\gamma = p_2' V_2'^\gamma \qquad (2)$$

Here unprimed quantities refer to the initial state, primed quantities to the final state, and $\gamma = C_p/C_v$, where $C_p$ and $C_v$ are the specific heats at constant pressure and volume, respectively.
The correct solution for the process as is actually occurs is obtained by imposing energy conservation

$$U = U_1 + U_2 = U_1' + U_2' \qquad (3)$$

Equality of the initial and final pressures on the two sides

$$p_1 = p_2, \quad p_1' = p_2' \qquad (4)$$

equality of the final temperatures on the two sides

$$T_1' = T_2' = T' \qquad (5)$$

and

$$V = V_1 + V_2 = V_1' + V_2' \qquad (6)$$

The final pressures on the two sides are given by

$$p'_1 = \frac{n_1}{V'_1} RT' \qquad (7)$$

and

$$p'_2 = \frac{n_2}{V'_2} RT' \qquad (8)$$

where $n_1$ and $n_2$ are the numbers of moles of the gas in the volumes $V'_1$ and $V'_2$ respectively.
From eqs. (4), (5), (6), (7) and (8) it follows that

$$V'_2 = \frac{n_2}{n_1 + n_2} V \qquad (9)$$

and

$$V'_1 = \frac{n_1}{n_1 + n_2} V \qquad (10)$$

The actual physical process is "quasi-static" (although not reversible), and since $U_i = U_i(V_i, S_i)$ $(i=1, 2)$

$$dU_1 = -p_1 dV_1 + T_1 dS_1 \qquad (11)$$

and

$$dU_2 = -p_2 dV_2 + T_2 dS_2 \qquad (12)$$

But, for two near points, where the piston kinetic energy is zero

$$dU_1 + dU_2 = dU = 0 \qquad (13)$$

Therefore, from eqs. (11)-(13) it follows that

$$-p_1 dV_1 + T_1 dS_1 = p_2 dV_2 - T_2 dS_2 \qquad (14)$$

Since the pressures on the two sides of the piston are equal at all times and the total volume is fixed (in fact it is possible demonstrate that the pressures are equal and constant during the stochastic movement to the final equilibrium point for every point where eq.(13) holds)

$$p_1 = p_2 \qquad (15)$$

and

$$dV_1 = -dV_2 \qquad (16)$$

It follows that for the actual physical process

$$T_1 dS_1 = -T_2 dS_2 \qquad (17)$$

Of course, for an actual physical process,

$$dS = dS_1 + dS_2 > 0 \qquad (18)$$

Equations (17) and (18) immediately lead to

$$dS_1 (1 - \frac{T_1}{T_2}) > 0 \qquad (19)$$

If $T_2 > T_1$, it follows that

$$dS_1 > 0 \qquad (20)$$

From eqs. (17) and (20) it follows that

$$dS_2 < 0 \qquad (21)$$

Moreover, if $T_2 > T_1$, then $dV_2 < 0$, $dU_2 < 0$ and $T_2\, dS_2 < 0$.

In this process just considered, the pressures on the two sides of the piston are equal at all times, which means no "work" is done. However, the energy transfer occurs through the agency of the moving piston, and if one considers "work" to be the energy transferred via macroscopic, non-random motion, then it appears that "work" is done. There is no paradox or problem here, since it is the net "work", the "work" done on side 1 by side 2 + the "work" done on side 2 by side 1, that is zero (see the generalization for the situation described in fig. 3). Furthermore, since the thermal conductivity of the piston is zero, no "heat" flow is possible. However, the flow of energy is from the high temperature side to the low temperature side, and if one considers heat to be energy transfer resulting from a temperature difference, then it appears that there is heat flow. Obviously, there is a problem as far as what we mean by heat and work [38-53]. An apparent solution of that problem, and this is the dominant orthodox point of view, is that heat is what complements work through the expression $dU=dW+dQ$ (clearly a tautological definition of heat). One problem is what is $dW$. Another is the physical meaning of quantity "heat", $dQ$. This physical meaning is extraneous to the formal definition and results from an abusive generalization to the adiabatic piston configuration (fig. 1) of the meaning for the same quantity in another configuration (fig. 2). Let's consider this other configuration, fig 2:

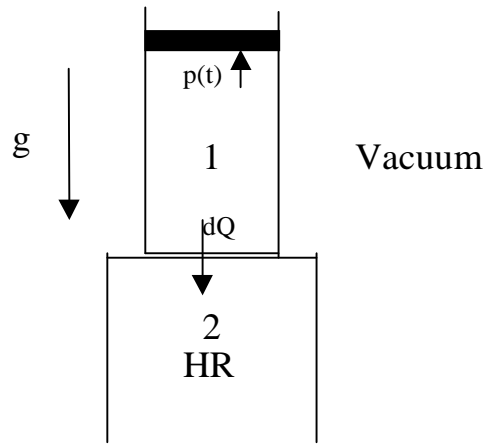

Fig. 2 - The sub-system 2, now, is a "Heat Reservoir". The quantity $dQ$ is the exchanged heat and is also the change of energy of the reservoir. Heat in this case as a fluid-like behavior.

When the piston moves, the instantaneous pressure $p(t)$ over the piston does the work $dW_p$. The weight of the piston does the work $dW_g$. The total work over the piston is $dW_t = dW_p + dW_g$. The kinetic energy of the piston is $E_k$ and the potential energy $E_p$. The internal energy of the gas, sub-system 1 is $U_1$ and the internal energy of the sub-system 2, the heat reservoir, $U_2$. The total work is equal to the kinetic energy change of the piston. We have from the energy conservation principle

$$dE_k + dE_p + dU_1 + dU_2 = 0 \qquad (22)$$

Since

$$dE_k = dW_t = dW_p + dW_g \qquad (23)$$

and

$$dE_p = -dW_g \qquad (24)$$

we have from (22),

$$dW_p + dU_1 + dU_2 = 0 \qquad (25)$$

We can write (25)

$$dU_1 = -dW_p - dU_2 \qquad (26)$$

or

$$dU = dW + dQ \qquad (27)$$

where $dU=dU_1$, $dW=-dW_p$ and $dQ=-dU_2$.

Consider now the configuration of fig. 3:

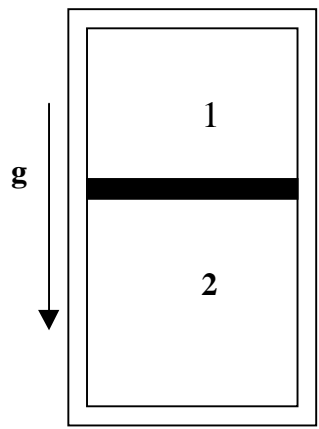

Fig. 3 - The adiabatic piston under gravity

Equations from (22) to (25) apply also for this configuration. Of course quantity $dW_p$ now means the total work of pressures on side 1 and 2 of the piston. Therefore from (25) we have

$$dW_{p_1} + dW_{p_2} + dU_1 + dU_2 = 0 \qquad (28)$$

or

$$dU_1 = dW_{p_1} + dW_{p_2} - dU_2 \qquad (29)$$

From (29), we have (27) but now $dW=dW_{p1}$ and $dQ= dW_{p2}-dU_2$

$$dU = dW + dQ \qquad (30)$$

with

$$dU = dU_1 \qquad (31)$$

$$dW = -dW_{p_1} \qquad (32)$$

and

$$dQ = -dW_{p_2} - dU_2 \qquad (33)$$

This quantity is not zero and therefore if we call this quantity heat we conclude that the piston is not adiabatic. Of course there is no paradox here because we are using the word heat in another sense. The energy exchanged between the two subsystems is due to the movement of the piston and not due to the energy exchanged through the piston without movement. Although the problem is probably not completely solved, rigorously solved, it is partially solved. It is not possible to continue to claim the impossibility of the movement of the adiabatic piston to the equilibrium with equal temperatures, if we admit a discontinuous interaction between the gas and the piston. Also it is possible to conceive a reversible transformation between two equilibrium points. For the reversible transformation the temperatures of the two gases are equal. However eq. (33) holds. Therefore, for this case, quantity $dQ$ has nothing to do with the energy exchanged between the gases as a consequence of a difference of temperature because there is no difference of temperature and quantity $dQ$ is not zero.

# Conclusion

The difficulties of the energetic interpretation based on the First Principle of Thermodynamics for the movement of an adiabatic piston submitted to the interaction of two gases that surround it are analyzed. The statement that a final equilibrium is achieved when the temperatures of the two gases are equal, has been denied for several years as a result of misinterpretations about the concept of heat. However, a kinetic and statistical analysis of the same problem gives the result of equal temperatures for equilibrium. This originates a well known controversy.

The problem presented here has a simple interpretation. It implies abandoning the point of view that heat is a special kind of energy exchanged between two bodies when a difference of temperature exists between them. In fact if we consider heat as internal energy (generalization of the kinetic interpretation of heat as the kinetic energy of atoms) it is possible to say that heat is "transformed" into work when a gas expands rising a piston. The mass associated with the piston has an increase of kinetic energy and potential energy. The work performed by the pressure over the piston between two points where the piston is stopped is equal in module to the work of the weight of the piston. And for a compression we can say that work is "transformed" into heat. Or if we consider the well known Joule's paddle wheel apparatus to demonstrate the equivalence of work and heat we have, with the internal energy interpretation of heat, a simple and direct interpretation of the experiment. As Joule did. If we consider two "heat reservoirs" at temperatures $T_1$ and $T_2$ we can say that heat "flows" from one reservoir two the other if the internal energy change of one reservoir is equal in module to the internal energy change of the other reservoir. And, if work is not zero, as when we consider a thermal machine, the work is equal to the internal energy changes of the reservoirs.

For the adiabatic piston considered we can say that heat flows (through the stochastic movement of the piston) from one gas to the other for the first configuration (fig.1) or that heat is transformed into work for the other configuration (fig. 3). For the last configuration a reversible transformation can be conceived when the temperatures are equal for the two sides of the piston. In this limiting case the jiggling of the piston between two equilibrium points exchanges energy between the two gases (without any difference of temperatures).

An other important point connected with the concept of heat is the asymmetry introduced by the Second Law: The internal energy is a function of the entropy, but the kinetic energy and the potential energy of the piston is not. When the deformation variables are the same, work is transformed into internal energy (heat), not the opposite. Some difficulties of interpretation of thermodynamics and in particular the adiabatic piston controversy result from several different concepts of heat. This can be no problem if we are aware of the non equivalence between this concepts [38-53]. But it can be a problem [7-37].